\documentclass[aps,preprint,showpacs,amssymb,superscriptaddress]{revtex4}
\usepackage{graphicx}
\usepackage{bm}
\usepackage{bm}
\begin{document}
\title{Critical Casimir Effect in $^3$He -$^4$He films}
\author{ A. Macio\l ek}
\affiliation{Institute of Physical Chemistry,
             Polish Academy of Sciences, Kasprzaka 44/52,
            PL-01-224 Warsaw, Poland}
\author{S. Dietrich}
\affiliation{ Max-Planck-Institut f{\"u}r Metallforschung, Heisenbergstr.~3, D-70569 Stuttgart, Germany}
\affiliation{
Institut f{\"u}r Theoretische und Angewandte Physik, Universit{\"a}t Stuttgart,
 Pfaffenwaldring 57, D-70569 Stuttgart, Germany}
\date{\today}
\begin{abstract}
Universal aspects of the thermodynamic Casimir effect in wetting
  films of  $^3$He-$^4$He mixtures
 near their bulk tricritical point  are  studied  within
suitable models serving as representatives of the corresponding 
universality class.
The  effective forces between the boundaries of such  films 
arising from the confinement are calculated along isotherms  at several 
fixed concentrations of $^3$He.
Nonsymmetric boundary conditions   
 impose  nontrivial concentration profiles leading to repulsive  
Casimir forces which  exhibit  a rich behavior of  the crossover 
 between the tricritical point and the line of critical points.
The theoretical results agree  with published experimental data and emphasize
the importance of logarithmic corrections.
\pacs{05.70.Jk, 64.60.Fr, 64.60.Kw, 67.40.Kh, 68.15.+e } 
\end{abstract}
\maketitle 
Finite-size contributions to the free energy of a fluid
confined between two surfaces at  a distance $L$  give rise 
to an effective  force between them.
 Theory predicts that at the
bulk critical point $T_c$ of such a system this force
becomes long-ranged as a result 
of critical fluctuations of the  corresponding ordering degrees of freedom.
This is analogous to the well-known Casimir effect
in electromagnetism. This so-called  critical 
Casimir force  $f_C$ per unit area and in units of $k_BT_c$ 
can be expressed in terms of  universal
 scaling functions~\cite{krech:99:0}.

Only recently, sophisticated  wetting experiments 
have provided detailed quantitative data for  critical Casimir forces
in various systems
~\cite{garcia:99:0,law:99:0,garcia:02:0,balibar:02:0,pershan}.
In the case of  $^4$He  wetting films near the
 superfluid transition, these experimental studies
support quantitatively theoretical predictions for
  $f_C$ ($T\ge T_c$) corresponding
 to the universality class
of the $XY$ model~\cite{krech:91:0}.
For the case of  $^3$He-$^4$He films 
near the  bulk  tricritical point some
theoretical predictions are  available \cite{krech:91:0},
but those do not apply for the boundary conditions
relevant for recent  wetting experiments performed in  these systems
~\cite{garcia:02:0,balibar:02:0}. 
However, the shape of the  scaling function of the Casimir force
 depends {\it sensitively} on
the type of boundary conditions (BC)  and thus 
on the   surface universality
classes to which the  confining  surfaces belong ~\cite{diehl:86:0}. 
The experiments of Ref.~\cite{garcia:02:0} report
a  {\it repulsive} $f_C$ around  the tricritical point
which  suggests {\it nonsymmetric} BC for
 the superfluid order parameter (SOP). This is  opposite to 
the case of pure $^4$He wetting films near the $\lambda$-point where 
  $f_C$ was found to be 
 {\it attractive}~\cite{garcia:99:0,krech:91:0}.
For the latter system the BC 
seem to be very well
 approximated by {\it symmetric}  Dirichlet-Dirichlet BC  $(O,O)$ forming 
 the so-called ordinary (O) surface universality class
 because the quantum-mechanical
 wave function describing
 the superfluid state vanishes at 
 both interfaces~\cite{krech:99:0,garcia:99:0}.
The type of BC for  $^3$He-$^4$He  wetting films
is not clear from the outset because 
 a $^4$He-rich layer forms near the substrate-fluid
 interface, which may become superfluid already above
 the bulk  $\lambda$-line~\cite{laheurte:78:0} whereas $^3$He has 
a preference for the fluid-vapor interface.
Thus the two interfaces impose a nontrivial concentration profile
which in turn couples to the SOP.
This leads to the hypothesis that the concentration profile
induces effectively nonsymmetric $(O,+)$ BC for the SOP, 
i.e., Dirichlet  boundary conditions at the fluid-vapor interface 
and symmetry-breaking (+) BC  at the substrate-fluid interface
(also known as the so-called extraordinary or normal universality class
\cite{diehl:86:0}). 
For the present tricritical behavior the upper critical dimension
$d^{\ast}$ equals 3.
In this case  theory predicts that for three-dimensional  systems 
the asymptotic tricritical thermodynamic functions  
exhibit  power laws with critical exponents taking their  classical values.
However,  logarithmic corrections to the mean-field  (MF)
behavior are expected  under experimental conditions~\cite{LawSar}.

Here we consider  two complementary approaches. 
Field-theoretical methods and renormalization-group (RG)
analyses are used to derive universal properties of the Casimir force
at the tricritical point and the form of 
logarithmic corrections. However, these methods  do not lend themselves
 for systematic studies
of $f_C$ along all thermodynamic paths followed in the
aforementioned experiments.  
In order to be able to interpret the  rich variation of  
$f_C$ extracted from the 
capacity measurements in Ref.~\cite{garcia:02:0},
 to understand the emergence of the actual BC
and, moreover, to predict the behavior of $f_C$ in the
crossover region   between the tricritical and the critical points,
we  employ the  vectoralized Blume-Emery-Griffiths model
 (VBEG)~\cite{maciolek:04:0} as a representative of the same universality
class as the actual physical system. This lattice model 
is extended to the  film geometry and treated within mean field theory (MFT).

First we derive the leading asymptotic behavior of 
$f_C$  at tricriticality for $(O,+)$ BC. 
To this end  we consider  the standard  Ginzburg-Landau  (GL)
Hamiltonian  for an $O(n)$-symmetric tricritical system ($T=T_t$)
 in a film geometry:
\begin{equation}
\label{eq:1}
{\cal H}[{\bf \Phi}]=\int d^{d-1}x\int_0^Ldz\left\{ \frac{1}{2}(\nabla {\bf \Phi})^2+\frac{u}{6!}({\bf\Phi}^2)^3 \right\}
\end{equation}
where $L$ is the film thickness, ${\bf \Phi}$ is the $n$-component OP and $z$
is the distance between the confining surfaces;
 $u$ is a  bare coupling constant. 
In a film geometry 
$f_C\equiv -(\partial f^{ex}/\partial L)=\langle {\cal T}_{zz}\rangle $
is given by the  stress tensor
component ${\cal T}_{zz}$~\cite{krech:99:0},
where $f^{ex}(L)\equiv (f-f_b)L$, $f$ 
is the total free energy per unit area  and per $k_BT_t$
and  $f_b$ is the bulk contribution.
The stress tensor is given by~\cite{krech:99:0}
${\cal T}_{ij}=\partial_i{\bf \Phi}\cdot \partial_j{\bf \Phi}-\delta_{ij}{\cal L}-(d-2)/(4(d-1))(\partial_i\partial_j-\delta_{ij}\nabla^2){\bf \Phi}^2$,
where ${\cal L}$ is the  integrand of (\ref{eq:1}).
We take ${\bf \Phi}=(m(z),0,\ldots,0)$. Determination of the 
tricritical Casimir force  starts
from the Euler-Lagrange equation for the OP  profile:
$m''(z)=(u/120)m^5(z)$ with  $(O,+)$ BC, i.e., 
$m(0) = 0 \quad \mbox{and} \quad m(L) = +\infty
$.
In this case the spatially constant $\langle {\cal T}_{zz}\rangle$  can be  
expressed as  $(1/2)(m'(0))^2$. 
With the scaling ansatz
$
m(z) = (u/360)^{-1/4} L^{-1/2} \varphi(z/L)$ and 
$\langle T_{zz}\rangle = (90/u)^{1/2} L^{-3}\Theta$,
and after integrating directly the first integral of the 
Euler-Lagrange equation
one  obtains 
$\Theta ^{1/3} =\displaystyle \int_0^{\infty}dp/\sqrt{1 + p^6}\simeq 1.40218$ 
by implementing the above BC.
Eventually, in units of $k_BT_t$ the MFT  result for the tricritical 
Casimir force $f_C^t$ in the case of $(O,+)$ BC is
$f_C^t=2.7568(4)\left( 90/u\right)^{1/2}L^{-3}$.
 Note that within MFT the parameter $u > 0$
remains undetermined. Its value follows from using
standard RG  arguments.
In  $d=3-\epsilon $ the above MFT result  yields
 the leading contribution in an $\epsilon$-expansion. After 
removing the uv singularity via renormalization 
the  asymptotic scaling behaviour of $f_C^t$  follows from substituting 
$u$ by the appropriate fixed-point  value
  $u^{\ast}\propto \epsilon$.
At $d=d^{\ast}$, and under spatial rescaling by a factor  $\ell$,
$u$ flows to its RG fixed point value
$u^* = 0 $ according to $\bar{u}(\ell)=(240 \pi^2)/\left((3n+22)|\ln \ell|\right)$  \cite{eisen:88}. With the  rescaling factor $\ell =l_0/L$, where $l_0$ is a microscopic length 
scale of the order of a  few \AA, ~this yields a logarithmic correction
 to the power law $L$-dependence of  the tricritical 
Casimir force:
\begin{equation}
\label{eq:as}
f_C^t\simeq 0.54(3n+22)^{1/2}(\ln L/l_0)^{1/2}L^{-3}.
\end{equation}
 Gaussian fluctuations 
give contributions of at least $O(u^0)$ which are therefore
subdominant. We compare Eq.~(\ref{eq:as}) for $n=2$ with 
the data obtained 
 by Garcia and Chan~\cite{garcia:02:0} for   their
experimental value   $L/l_0\approx$ 520 \AA /1.3 \AA. This  gives
$
\vartheta_t\equiv f_C^t L^{3} \approx 6.96$
in a good  agreement with 
 $\vartheta_t^{exp}=8.4\pm 1.7$, which  suggests that this experiment 
maybe the first to have verified implicitly the existence 
of logarithmic corrections
near the tricritical point. 
However, in order to  extract the actual
value of the {\it universal} Casimir amplitude (i.e., the numerical prefactor in Eq.~(\ref{eq:as})) the experimental data require a reanalysis
based on a functional form given by Eq.~(\ref{eq:as}).

\begin{figure}
\begin{center}
\includegraphics[width=0.9\textwidth]{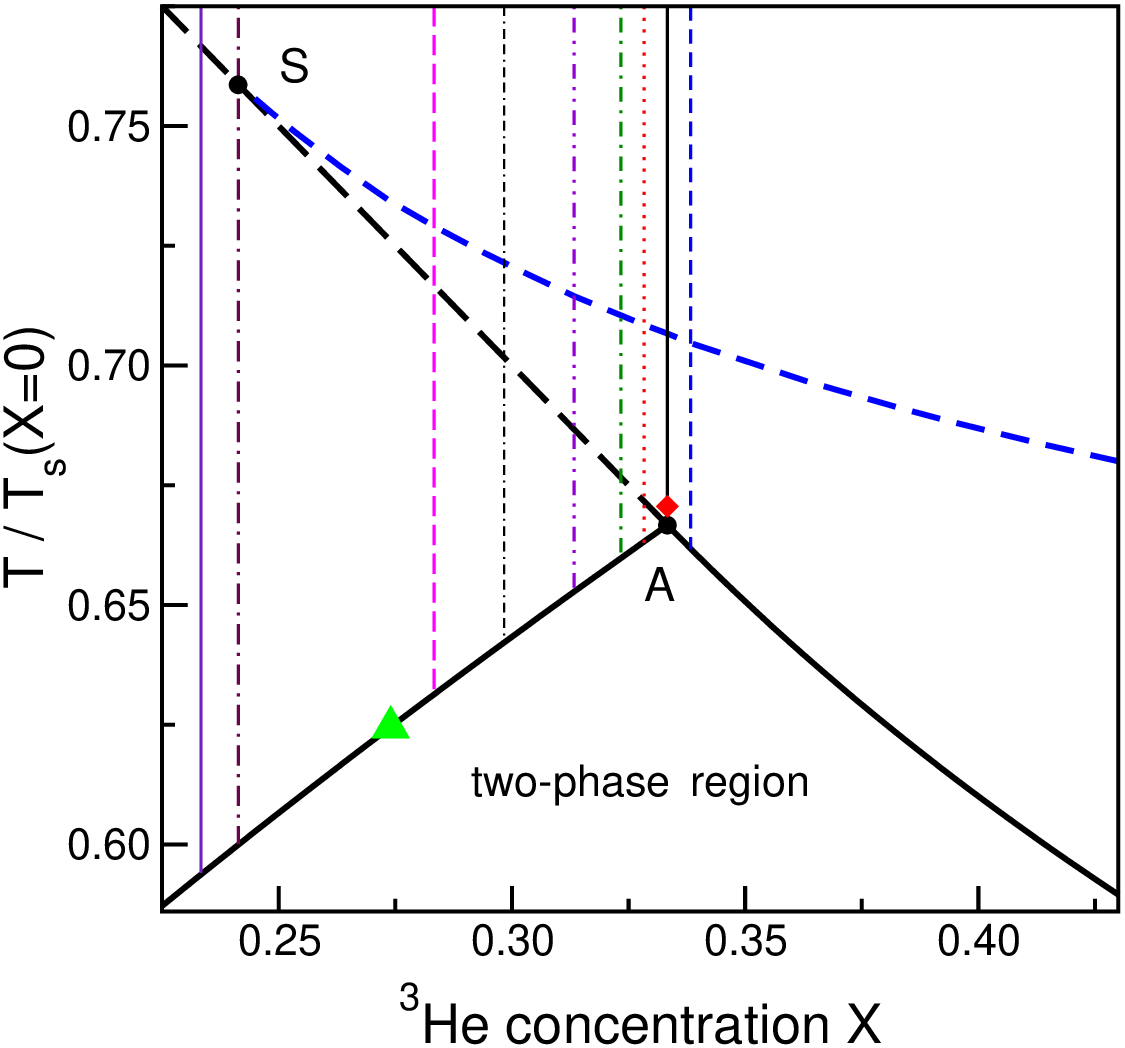}
\end{center}
\vskip15pt
\caption{ 
Phase diagram for the VBEG model 
obtained whithin MFT
  for $K/J=0.5$ and $\Delta _1/J=-3$ exhibiting the bulk 
$\lambda$-line $T_s(X)$ of continuous 
 superfluid transitions  (long-dashed line),
 the  phase separation curves (solid lines), 
the tricritical point  $A=(T/T_s(0)=2/3,X=1/3)$, and the
 surface transition line (short-dashed line) which merges with the bulk
$\lambda$-line at the special transition point $S=(T/T_s(0)\simeq 0.759, X\simeq 0.241$).  Vertical lines represent thermodynamic paths along which
the Casimir force  has been calculated (see Fig.~\ref{fig:3}). $ \blacklozenge$,$\bullet $ (A), $\blacktriangle$: state points considered in Fig.~\ref{fig:2}.}
\label{fig:1}
\end{figure}

\begin{figure}
\vskip15pt
\begin{center}
\includegraphics[width=0.9\textwidth]{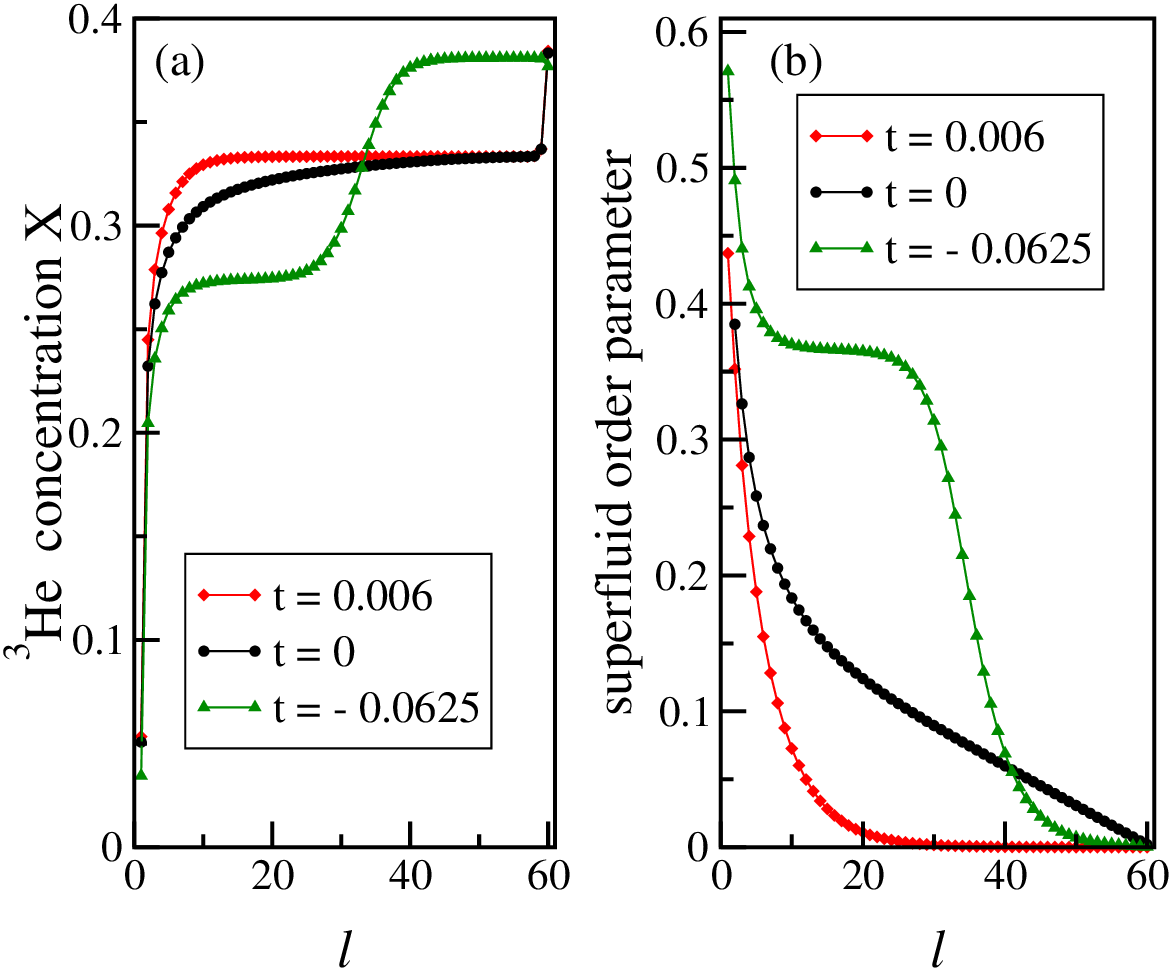}
\end{center}
\vskip15pt
\caption{(a) $^3$He concentration profile $X(l)=1-Q_l$   and (b) SOP 
$m_l$ profile  for a VBEG film of  width
$L=60$ for  $K=0.5J$, 
$\Delta_1/ J=-3$, and $\Delta_2/ J=\Delta_t/ J\simeq 0.61$
 corresponding to the state points $\blacklozenge$, $\bullet$, and 
$\blacktriangle$ indicated in Fig.~\ref{fig:1}.
}
\label{fig:2}
\end{figure}

\begin{figure}
\begin{center}
\includegraphics[width=0.9\textwidth]{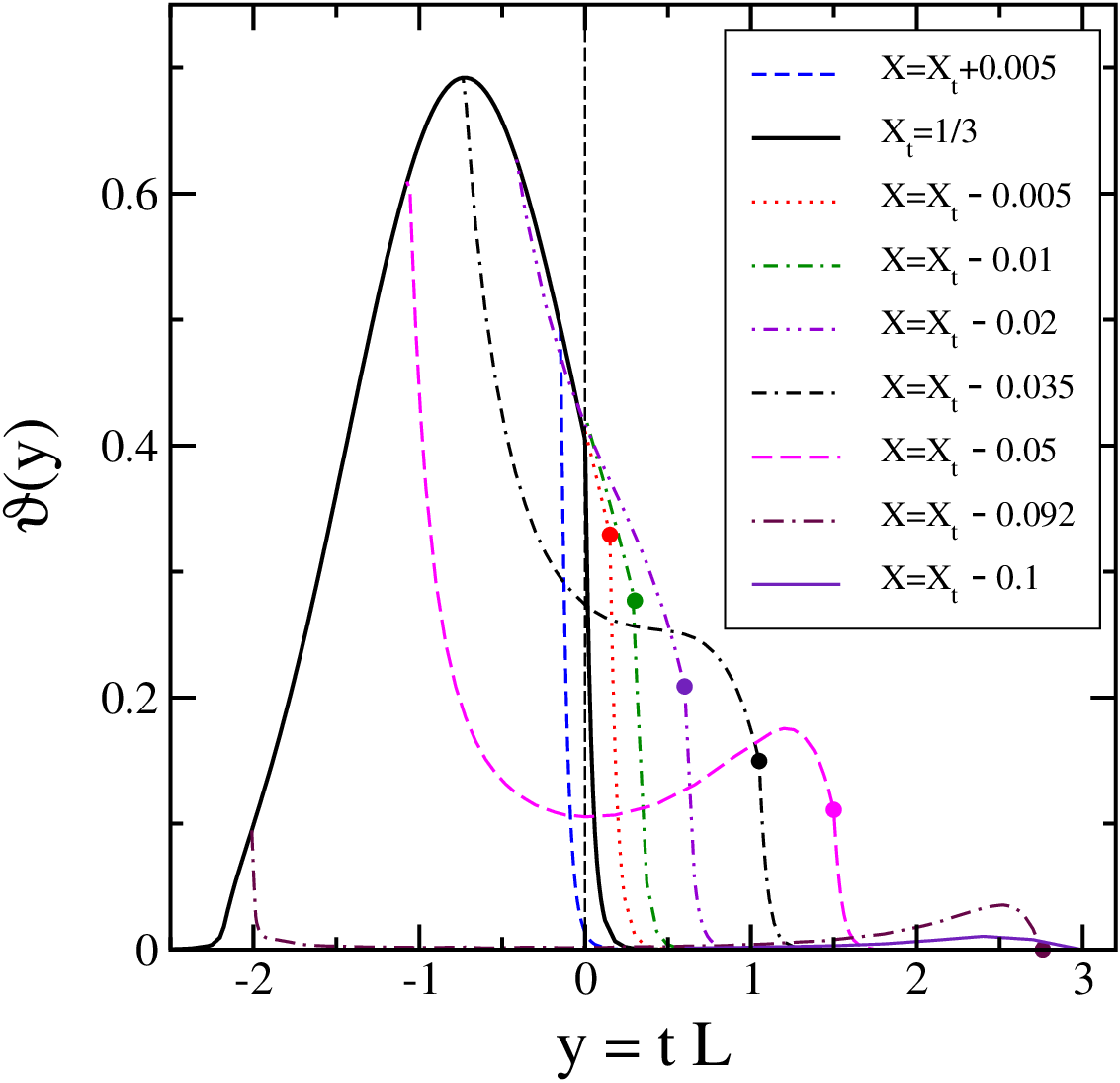}
\end{center}
\vskip15pt
\caption{Color online. Dimensionless scaling function 
${\vartheta}(y=tL)=f_CL^3$ 
 for the VBEG model calculated   within  MF theory
along the paths of fixed  concentration of $^3$He shown in Fig.~1.
Dots indicate the corresponding onset temperature $T_s(y)$ of  superfluidity
at the $\lambda$-line.  }
\label{fig:3}
\end{figure}

\begin{figure}
\vskip15pt
\begin{center}
\includegraphics[width=0.9\textwidth]{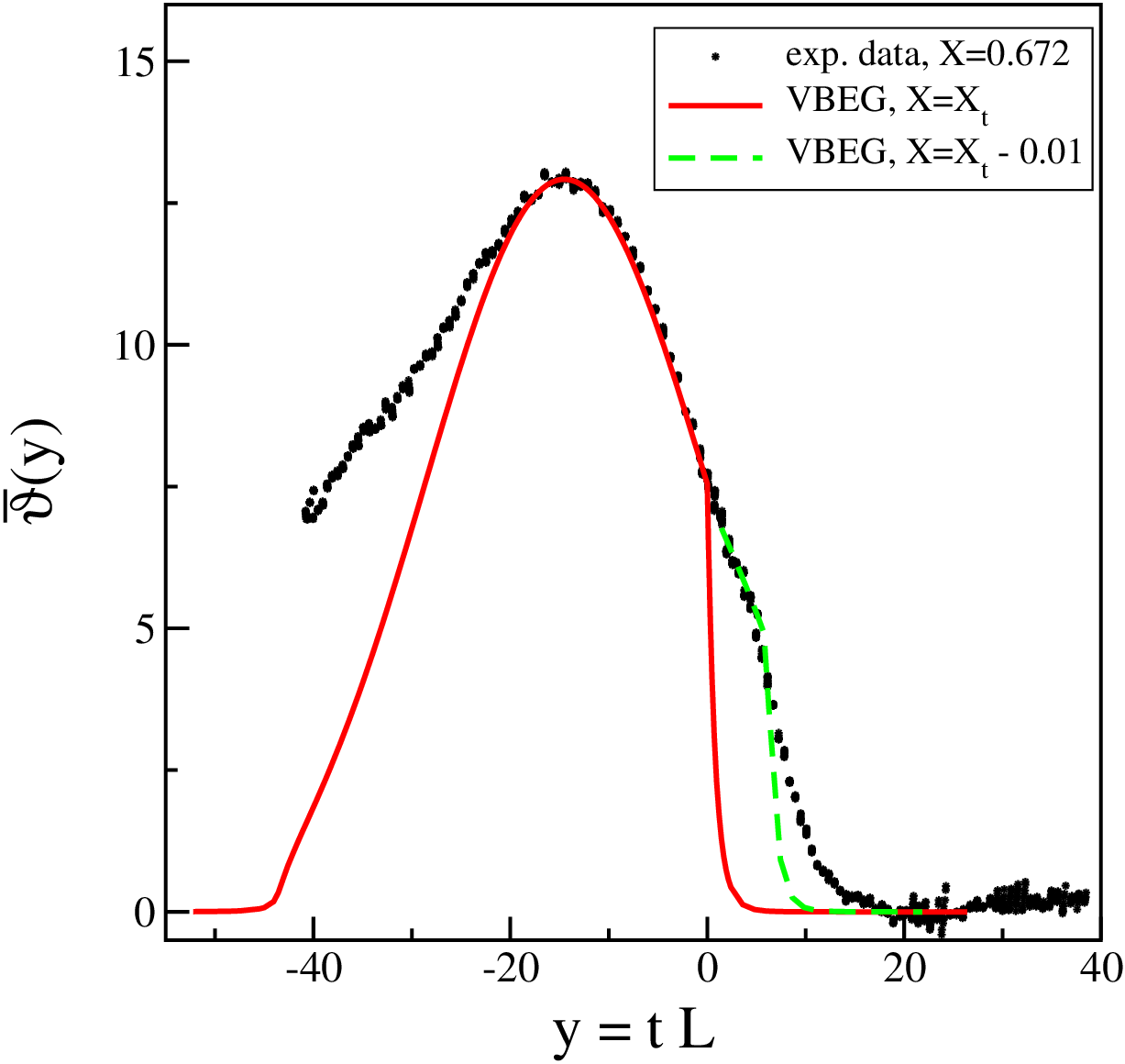}
\end{center}
\vskip15pt
\caption{ The  adjusted scaling function ${\bar \vartheta}(y)$ 
(see main text)
 for the  VBEG model within  MFT compared with the corresponding  
experimental curve \cite{garcia:02:0} obtained along the path of
 fixed tricritical
 concentration $X_t\approx 0.672$ of $^3$He. }
\label{fig:4}
\end{figure}

Now we turn to the VBEG model and consider  a $d=3$  simple 
cubic lattice consisting of $L$ parallel  lattice layers at spacing $a$.
 Each layer has $A$ sites,
labeled $i, j, \ldots$ and  associated with an occupation  variable  $t_i=0, 1$
and a phase $\theta _i$ $( 0\le \theta _i< 2\pi)$ which mimics
 the phase of the $^4$He  wave function.
A $^3$He ($^4$He) atom at site $i$ corresponds  to $t_i=0 (1)$.
 The Hamiltonian is given by 
\begin{equation}
\label{eq:ham}
{ \cal H}=-J\sum _{<ij>}t_it_j\cos (\theta _i-\theta _j) -K\sum _{<ij>}t_it_j+ \sum _i\Delta_i t_i
\end{equation}
where    the first two sums run over  nearest-neighbor pairs and the
last one  is over all lattice sites.
The  field $\Delta _i$ is related to the  chemical potentials
 of the two  components of the mixture.
 $\Delta_i=\Delta_1$ and $\Delta_i=\Delta_2$ on the left and right surface
layer, respectively, and $\Delta_i=\Delta $ otherwise.
 The differences
 $\Delta_i-\Delta,~ {i=1,2}$, are a measure of   the relative preferences
of $^4$He atoms for the two  surfaces such that $\Delta _1<\Delta$ corresponds
to the preference of $^4$He atoms for the 
solid substrate.
At  the opposite surface with the vapor we choose $\Delta _2=\Delta _t$, 
the tricritical bulk  value. Near the tricritical point this  choice 
 is consistent with the assumption made in Ref.~\cite{garcia:02:0} 
for the concentration profile across  the wetting film, whereby at 
the interface with the vapor the 
 $^3$He mole fraction takes the  bulk value.
 Bulk properties of  this model  were studied within
 MFT  and by Monte Carlo simulations 
in $d=3$  in \cite{maciolek:04:0}. The resulting bulk phase diagram 
resembles that observed  experimentally  for
$^3$He-$^4$He mixtures (see Fig.~\ref{fig:1}).
For the film geometry we have solved this model
within  MFT. This  yields
a  set of self-consistent 
 equations for the OP in the $l$th layer,
i.e., the concentration  $Q_l=<t_l>=1-X(l)$ of  $^4$He,  and 
the SOP ${\bf M}_l=(m^{(1)}_l,m^{(2)}_l)$ with
$m^{(1)}_l=\langle t_{l}\cos \theta _{l}\rangle$ and 
 $m^{(2)}_l=\langle t_{l}\sin \theta _{l}\rangle$
where $t_l$ and $\theta _l$ denote the occupation
number and the phase in the $l$th layer, respectively.
In the absence of helicity one has  ${\bf M}_l=(m^{(1)}_l,0)\equiv (m_l,0)$,
$Q_l=I_0({\tilde J}b_l)/\left(e^{-{\tilde K}a_l+\Delta_l/T}+I_0({\tilde J}b_l)\right) $,
and
$
m_l=I_1({\tilde J} b_l)/\left( e^{-{\tilde K}a_l+\Delta_l/T}+I_0({\tilde J}b_l)\right)$.
$I_0(z)$ and  $I_1(z)$ are  modified Bessel functions and
$\Delta_l=\Delta$ for $l\ne 1,L$; 
${\tilde K}=qK/T$ and ${\tilde J}=qJ/T$, where  $q=q_{\perp}+2q'$ is  the coordination number of the lattice. $q_{\perp}$ is the in-layer coordination 
number while each site  (but not in the first and last layers) is connected 
to $q'$ atoms in each  adjacent layer. 
We have introduced  $b_l\equiv m_{l-1}+q_{\perp}m_l+m_{l+1}$ for $l\ne 1, L$,
 $b_1\equiv q_{\perp}m_1+m_2$, and  $b_L\equiv m_{L-1}+q_{\perp}m_L$
and analogously
  $a_l\equiv Q_{l-1}+q_{\perp}Q_l+Q_{l+1}$ for $l\ne 1, L$,  $a_1=q_{\perp}Q_1+Q_2$, and $a_L=Q_{L-1}+q_{\perp}Q_L$.
The coupled equations for $Q_l$ and $m_l$ are solved numerically; 
the acceptable
solution minimizes the free energy.
First,  we have analyzed
 the semi-infinite system. Close to the $\lambda$-line  we observe a
 higher $^4$He concentration near the left   surface,
 which  induces   a local superfluid ordering (see Fig.~\ref{fig:2}).
 By varying $T$ and
$\Delta$ one obtains a whole  line of continuous
 surface transitions  corresponding to the onset of the formation
of  a superfluid film near the wall; it
meets the $\lambda$-line  
at the special transition point whose position depends on the value
of    $\Delta _1$ (see Fig.~\ref{fig:1}). These  findings are in  
 agreement with the results
 of a Migdal-Kadanoff analysis~\cite{peliti:85:0}.
In the film geometry the Casimir force is obtained by taking 
a finite difference
after  calculating $f^{ex}$ for  $L_0$ and
$L_0+1$.
Figure~\ref{fig:3} summarizes our result for a  film of width  $L=20$,
$K/J=0.5$, $\Delta _1/ J=-3$, and 
$\Delta _2=\Delta _t/J\simeq 0.61$. $f_C$ is calculated along the
 thermodynamic  paths indicated in Fig.~\ref{fig:1}.
Below $T_t$,  $f_{C}$ is calculated along the coexistence 
line, infinitesimally on the superfluid branch of  bulk coexistence.
Our results are presented in terms 
of the  scaling function 
 $\vartheta \equiv L^d f_C$ as a function of the scaling variable 
 $y\equiv tL^{1/\nu}=(L\xi_0/\xi)^{1/\nu}$, where  $t=(T-T_t)/T_t$.
$\xi_0$ is the amplitude of the correlation length $\xi$ and $\nu =1$.
The surface transition does not leave a visible trace 
in the behavior of $f_{C}$.
 For $^3$He concentration  $X < X_t$, upon crossing the $\lambda$-line there is a steep
variation associated with a break in slope, giving rise to the formation
of  shoulders which are similar to those observed 
experimentally \cite{garcia:02:0}. For $X > X_t$, when   $T$  reaches the
  phase separation temperature $f_{C}$ coincides with
 the curve common to all  values of $X$. This occurs 
 with a discontinuous  first  derivative.
 The aforementioned common curve exhibits a pronounced maximum 
below $T_t$ at 
$y\simeq -0.74$ and gradually  decreases to zero for
 $y\to -\infty$.
 Below $T_t$, both the concentration  and the SOP  
profiles corresponding to this common curve display
 an interface-like structure  separating two 
domains of the coexisting bulk phases
(see $t=-0.0625$ in  Fig.~\ref{fig:2}).
Features of $f_{C}$ in this  'soft mode' phase can be attributed to
purely interfacial effects, similarly to  Ising-like
 films with asymmetric BC~\cite{parry:92:0}.  Beyond MFT a
 positive sign of  the force can be regarded
as a consequence of entropic repulsion~\cite{fisher:86:0}.
The maximum of $f_C$ is expected to
 occur at the temperature $T$ for which  the interfacial width
$\sim \xi _b\sim L$, i.e., $y\simeq -1$~\cite{parry:92:0} which checks
 with Fig.~\ref{fig:3} \cite{footnote}.
For $X \le X_t-0.05$ we observe a  crossover to the critical superfluid
behavior of pure $^4$He and a  gradual formation of a second, less pronounced 
local maximum located slightly below the $\lambda$-line.
 This local maximum decreases 
upon departure from $X_t$ and finally disappears
above the special transition $S$. This is expected,
because  above $S$ the BC turn into the type $(O,O)$ for which 
$f_{C}$ vanishes within MFT. 
For lower $T$, $f_{C}$ increases steeply upon  approaching bulk coexistence
revealing that   interfacial effects associated with the 
'soft mode'  lead  to a much stronger Casimir effect than near the critical 
$\lambda$-line.
The qualitative features of   $\vartheta$
extracted from the experimental data for $X\simeq X_t$
(see Fig.~5 in  Ref.~\cite{garcia:02:0})
are very  well captured by the present  lattice model.
Discrepancies can be attributed to  fluctuation
effects neglected in the present MFT VBEG approach:
(i) The discontinuities of slopes as obtained within MFT
upon  crossing the $\lambda$-line are expected to be  smeared out 
by fluctuations.
(ii) The experimental scaling function $\vartheta$ does not vanish  
at low temperatures, which may be due to  Goldstone modes 
in the superfluid phase. However,
the possibility that this behavior is an artifact of an extreme change
in the dielectric constant of the  film cannot
 be excluded~\cite{privat}. 
In the crossover regime to the critical behavior 
only few experimental  data for the thickness of the wetting 
film are published. However,
again the  variation of film thicknesses agrees with our findings.
In particular, one observes a rapid thickening of the films
upon  approaching the coexistence line; for some values of  $X$ a small
 maximum 
located near  the $\lambda$-line is also visible (compare Fig.~\ref{fig:3}).
A  quantitative comparison is not possible because, 
 for our choice of surface terms in the  Hamiltonian, 
the  fixed-point BC $(O,+)$ cannot be reached within the VBEG model.
 In order to be able to extract universal
properties - which requires to reach fixed-point BC - 
 it would be necessary to introduce a 
 surface field which couples to the SOP so that the BC (+) 
 can be realized. Also MFT is not sufficient, even in $d=d^{\ast}$. 
A naive correction of  $\vartheta$ by the logarithmic 
factor derived within the GL  model will not give 
the proper universal behavior. Instead, renormalization group schemes 
 or  Monte Carlo simulations have to be employed.
Nonetheless,  our MFT results for $X=X_t$, 
if  matched at the tricritical point $y=0$
and after adjusting the amplitude $\xi _0$ so that the positions
 of the maximum
of the scaling function are the same (i.e.,  
 $\xi^{th}_0/a\approx 0.065$)
reproduce  very well  the experimental curve (see Fig.~\ref{fig:4}),
 especially near the maximum where we expect interfacial
effects to be dominant. This is consistent because the 
'soft mode' phase does not depend on the details of the surface fields.  
Notice, that the experimental data nominally
 for $X=X_t$ more closely match the theory for
$X=X_t-0.01$. This raises the question as to whether the $^3$He
concentration in the film is shifted  relative 
to the bulk one.
 
A.M. benefited from discussions with  R. Garcia, M. Krech
 and S. Kondrat.
This work was partially funded by KBN grant No.4 T09A 066 22.

\vfill \eject
\vfill \eject

\vfill \eject

\vfill \eject

\begin{thebibliography}{99}
\bibitem{krech:99:0} M. Krech, {\it The Casimir Effect in  Critical System}   
(World Scientific, Singapore, 1994); J. Phys. Condens. Matter {\bf 11}, R391 (1999).

\bibitem{garcia:99:0} R. Garcia and M. H. W. Chan, Phys. Rev. Lett. {\bf 83}, 1187 (1999).


\bibitem{law:99:0}A. Mukhopadhyay and B. M. Law, Phys.~Rev.~Lett. {\bf 83}, 772 (1999).


\bibitem{garcia:02:0} R. Garcia and M. H. W. Chan, Phys. Rev. Lett. {\bf 88}, 086101 (2002).

\bibitem{balibar:02:0} T. Ueno, S. Balibar, T. Mizusaki, F. Caupin, and
E. Rolley, Phys. Rev. Lett. {\bf 90}, 116102 (2003).

\bibitem{pershan} M. Fukuto, Y. F. Yano, and P. S. Pershan, Phys. Rev. Lett. {\bf 94}, 135702 (2005).

\bibitem{krech:91:0} M. Krech and S. Dietrich, Phys. Rev. Lett. {\bf 66}, 345 (1991); {\bf 67}, 1055 (1991); Phys. Rev. A {\bf 46}, 1886 (1992); {\bf 46} 1922 (1992).

\bibitem{diehl:86:0} H. W. Diehl, in {\em Phase Transitions and Critical
Phenomena}, edited by C. Domb and J. L. Lebowitz (Academic, London,
1986), Vol. 10, p.76.

\bibitem{laheurte:78:0} J.-P. Romagnan, J.-P. Laheurte, J. -C. Noiray, and W. F. Saam, J. Low Temp. Phys.~{\bf 30}, 425 (1978).

\bibitem{LawSar}
D. Lawrie and S. Sarbach, in {\em Phase Transitions and Critical
Phenomena}, edited by C. Domb and J. L. Lebowitz (Academic, London,
1984), Vol. 9, p.2.\bibitem{eisen:88} E. Eisenriegler and H. W. Diehl, Phys. Rev. B, {\bf 37}, 5257 (1988).

\bibitem{maciolek:04:0} A. Macio\l ek, M. Krech, and S. Dietrich, Phys. Rev. E
{\bf 69}, 036117 (2004); and references therein.

\bibitem{peliti:85:0} A. Crisanti and L. Peliti, J. Phys. A: Math. Gen.
 {\bf 18} L543 (1985).

\bibitem{parry:92:0} A. O. Parry and R. Evans, Phys. Rev. Lett. {\bf 64}, 439 (1990).

\bibitem{fisher:86:0} M. E. Fisher, J. Chem. Soc. Faraday Trans. II {\bf 82}, 1569 (1986).
\bibitem{footnote} The  experimental data for $f_{C}$  exhibit a maximum
at $y\simeq -18$ does not relate to the condition $\xi _b \sim L$. 
\bibitem{privat} R. Garcia, private communication. 
\end{thebibliography}
\end{document}